\documentclass{article}
\usepackage[T1]{fontenc} 
\usepackage[utf8]{inputenc} 
\usepackage{ismir,amsmath,cite,url}
\usepackage{graphicx}
\usepackage{booktabs}
\usepackage{color}
\usepackage{soul}
\usepackage{subcaption}

\title{Artist and style exposure bias in collaborative filtering based music recommendations} 

\twoauthors
  {Andres Ferraro, Dmitry Bogdanov, Xavier Serra} {Music Technology Group - \\ Universitat Pompeu Fabra \\ {\tt first.lastname@upf.edu}}
  {Jason Yoon} {Kakao Corp. \\ \tt{jason.yoon@kakaocorp.com}}

\begin{document}

\maketitle
\begin{abstract}
Algorithms have an increasing influence on the music that we consume and understanding their behavior is fundamental to make sure they give a fair exposure 
to all artists across different styles. 
In this on-going work we contribute to this research direction analyzing the impact of collaborative filtering recommendations from the perspective of artist and music style exposure given by the system. 
We first analyze the distribution of the recommendations considering the exposure of different styles or genres and compare it to the users' listening behavior. 
This comparison suggests that the system is reinforcing the popularity of the items. Then, we simulate the effect of the system in the long term with a feedback loop. From this simulation we can see how the system gives less opportunity to the majority of artists, concentrating the users on fewer items. 
The results of our analysis demonstrate the need for a better evaluation methodology for current music recommendation algorithms, not only limited to user-focused relevance metrics.

\end{abstract}

\section{Introduction}

There are multiple factors that make design decisions for a music recommendation system a complex problem.
Some decisions can be related to theoretical aspects of music,  while others may have ideological or social connotations, may be subjective, not possible to quantify, or be changing depending on time and context~\cite{schedl2015music}.

Collaborative filtering methods are typically used to generate a recommendation by identifying patterns in what people listen from historical information. 
The drawback of these methods is that since they do not consider any other than information about interactions between users and items, it is not possible to generate recommendations for new items (the cold-start problem). Also the recommendations tend to follow the distribution of popularity of the music~\cite{knees2013survey} with the most popular items being recommended more (the long-tail recommendation problem). 
Celma and Cano~\cite{celma2008hits} show this 
by analysing navigation, clustering and connectivity in artist similarity networks built with collaborative filtering data.


With the advances in deep learning, new methods had been proposed for long-tail and cold-start recommendations using audio information and metadata~\cite{oramas2017deep,van2013deep}, which can learn automatically a representation from the data without the need for manually selecting the features.

Still, these solutions have some issues, in particular related to the fact that they work as black-boxes. For example, it is difficult to explain the results and it is hard to know if different musical styles are well-represented. Also, previous works do not show 
how robust these methods are to biased datasets 
and if it is possible to generate recommendations for new styles or genres that are less present in the user-item interactions. 

The growth of music streaming services in the last years has increased the importance of music recommender systems, and reducing the choice overload is commonly referred to as one of the advantages of these systems. 
Therefore, it is important to understand the increasing impact that these systems have to what people listen. They define which song will be the next hit, how much will an artist earn or even which music genres might receive 
almost zero promotion. 
This raises some ethical issues that had been discussed in previous works. For example, Holzapfel et al.~\cite{holzapfel2018ethical} raise the question if a group of artists that are never recommended by a system can be considered a case of discrimination. As researchers, we have to think about the implications of the systems we develop and the importance of assuring every artist has a fair chance to reach the public~\cite{ferraro2019music}. 

Recently, there have been studies trying to address these issues. Cramer et al.~\cite{cramer2019translation} summarizes 
possible algorithmic biases 
and highlights that music recommendations for ``balanced'' not-biased consumption may not necessarily lead to optimal experience for many users. 
McInerney et al.~\cite{mcinerney2018explore} propose a bandits approach to balance exploration and exploitation in the recommendations for the users, but they do not address its impact on the exposure of different artists or music styles. Mehrotra et al.~\cite{mehrotra2018towards} proposes a way to understand the trade-off between relevance, satisfaction and fairness in music recommendations. In this case, fairness measures the diversity of the level of popularity of recommendations, but it does not capture the overall exposure of the artists or the different musical styles.

Following these studies, 
we demonstrate preliminary results of our on-going research 
that gives a better understanding of the influence of music recommendation systems on users' behavior that could affect artists' exposure. We show that the distribution of the recommendations 
in terms of their artists, styles or genres is different from what the users had listened before. Also, we show that 
with time the system  tends to recommend fewer items, therefore, focusing user interactions on fewer artists, which is not the desired behavior of the system.

\section{Proposed analysis}\label{results}

In this work, we use a basic Matrix Factorization \cite{hu2008collaborative} algorithm and \textit{Echo Nest Profile Subset} to build a user-track matrix and generate 10 track recommendations for each user. 
We use the associated tags from \textit{Last.fm Dataset} to analyze how recommendations are distributed across the different musical styles in comparison with listening statistics from our dataset representing the initial preferences of users. 
We also show a simulation of how these recommendations can affect user 
behavior in the long term. For this we take the recommendation of the system for each user and increase the counter in the original user-track matrix, simulating that the users listened to all recommendations by the system. 
We then retrain the model and generate new recommendations. We repeat this process 30 times. 


\subsection{Datasets}\label{datasets}

The Million Song Dataset (MSD) \cite{mcfee2012million} is a large dataset of audio features and metadata expanded by the Music Information Retrieval community with additional information including tags, lyrics and other annotations. The Echo Nest Taste Profile Subset \cite{Bertin-Mahieux2011} provides play counts by 1,019,318 users covering 384,546 songs from MSD, originally gathered from an undisclosed set of applications. For this work we only consider users and items with more than 30 interactions (128,374 tracks by 18,063 artists and 445,067 users), to make sure we have enough information for training and evaluating the model. Additionally, the Last.fm Dataset \cite{Bertin-Mahieux2011} provides song-level tags extracted from \textit{Last.fm} for a subset of MSD. These tags are crowdsourced and cover genre, instrumentation, moods and eras. One track can have multiple tags.

\subsection{Metrics}


For a better understanding of system behavior, 
we need to define 
metrics that can assess how probable it is for new or less popular artists to be recommended and 
compare those across different styles. It is also valuable 
to know to how many different users each artist is recommended. 

In this work, we use the \textit{Gini index} to measure the distribution of how 
many users each artist gets recommended to,
but in future works other metrics should be also considered 
(for example, proposed for multistakeholder recommendation approaches~\cite{abdollahpouri2019beyond}).



We also use \textit{Coverage} to measure the percentage of different artists globally recommended. With this metric we can have an idea of the amount of artists that the system gives zero promotion.

\section{Results}\label{results}

\subsection{Distribution of recommendations}

\begin{figure}
 \centerline{
 \includegraphics[width=\columnwidth]{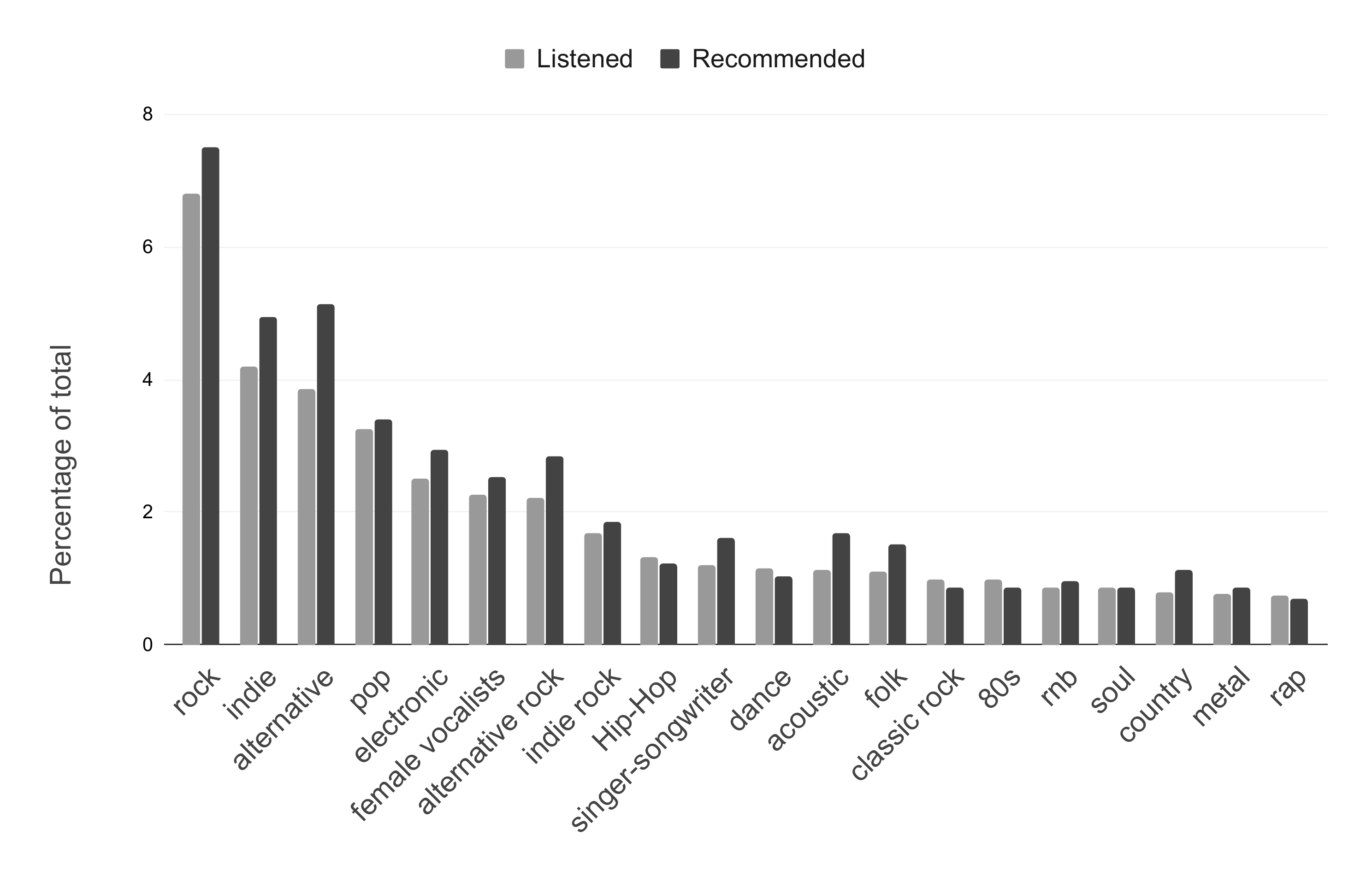}}
 \caption{Distribution of recommendations and users listening. Values are average percentages per music style.}
 \label{fig:dist}
\end{figure}

Figure \ref{fig:dist} shows the global tag distribution of all user-track recommendations pairs (10 tracks per user) compared to such a distribution for initial user listening behavior for the top 20 tags.\footnote{Note that there are 51,699 tags and therefore it is not possible to show all of them.} For the rest of the tags, the system is recommending 9.4\% less compared with what the users listened to.
Table~\ref{tab:rec-listen} similarly reports an average percentage of recommendations and initial user preferences in three tag and artists categories grouped by their popularity in terms of the original play counts.

We can see a clear popularity bias in what users listen to, 
and this bias is further reinforced by recommendations, which may be not the desired behavior. 
The system is recommending more top tags and less long-tail tags than what people listened to. 

\begin{table}[!h]
\footnotesize
 \begin{subtable}{\linewidth}
  \centering
 \begin{tabular}{lccc}
  \toprule
  & \textbf{Tags 1-5} & \textbf{Tags 5-2k} & \textbf{Tag 2k-50k}\\
  \midrule
 
Recommended & \textbf{4.7807} & \textbf{0.0347} & 0.0001 \\
Listened & 4.1195 & 0.0327 & \textbf{0.0003} \\
\bottomrule
 \end{tabular}
 \caption{Tags}
 \end{subtable}
 \begin{subtable}{\linewidth}
 \centering
 \begin{tabular}{lccc}
  \toprule
  & \textbf{Artists 1-5} & \textbf{Artists 5-2k} & \textbf{Artists 2k-18k}\\
  \midrule
 Recommended & \textbf{1.5672} & \textbf{0.0433} & 0.0003 \\
Listened & 0.6182 & 0.0370 & \textbf{0.0014} \\
\bottomrule
 \end{tabular}
 \caption{Artists}
  \end{subtable}
 \caption{Average percentage of recommendations and user play counts for (a) tags and (b) artists with different popularity. 
 }
 \label{tab:rec-listen}
\end{table}

\subsection{Simulating feedback loops}
\begin{figure}
 \centerline{
 \includegraphics[width=\columnwidth]{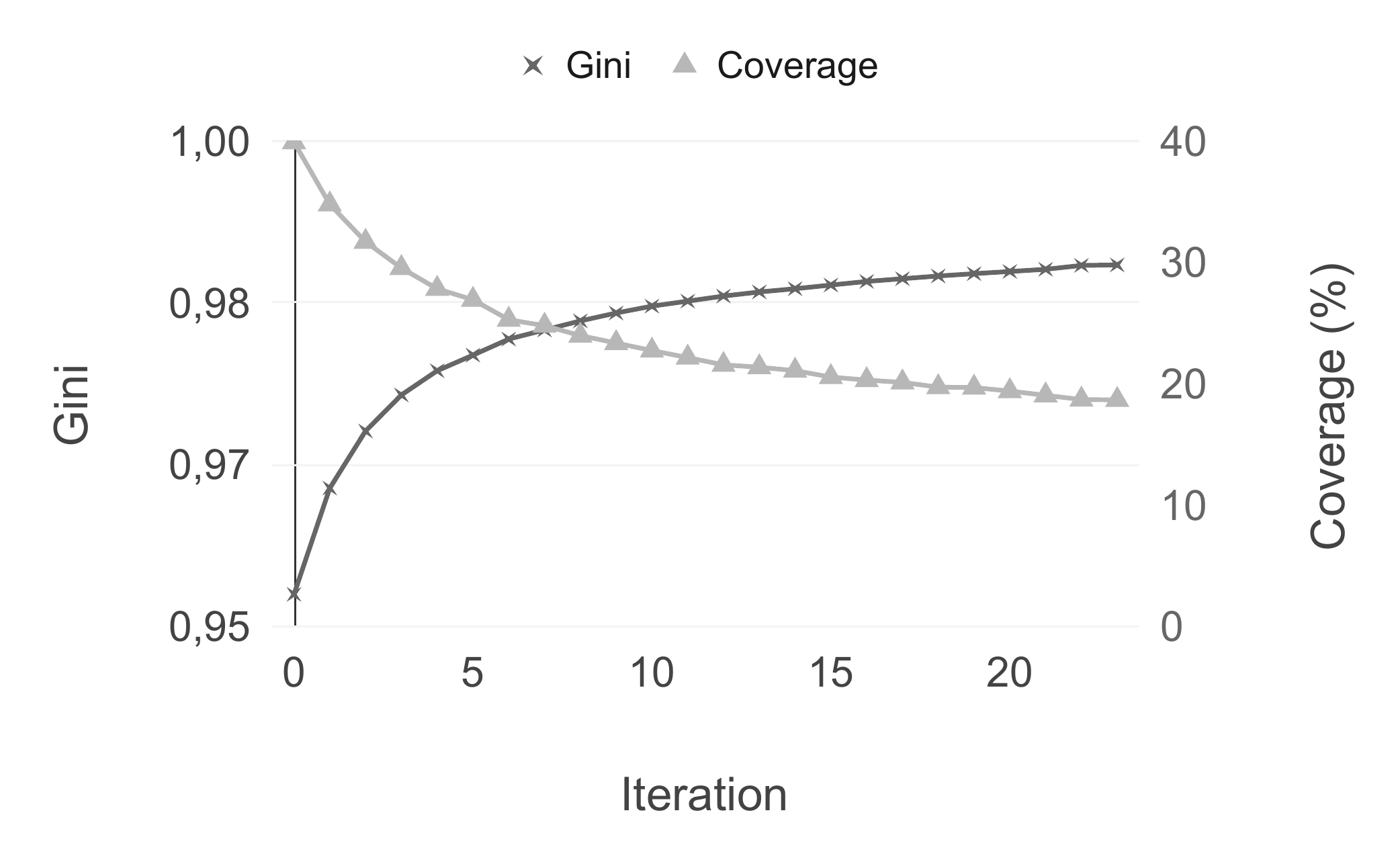}}
 \caption{Coverage and Gini index of the recommendations simulating feedback loops.}
 \label{fig:all}
\end{figure}

Figure \ref{fig:all} shows the results of simulating the feedback loop of the recommendations. We can see how the \textit{Gini index} increases on each iteration, starting in 0.95 and going up to 0.98. A value of 1.0 indicates that the system is recommending the same songs to all users. In the same figure we see the evolution of the \textit{Coverage} of the recommendations. For the first iteration the \textit{Coverage} is 40 \% but at the last iteration the \textit{Coverage} is 20 \% meaning that 80 \% of the songs are not recommended by the system.

\begin{figure}
\footnotesize
 \centerline{
 \includegraphics[width=\columnwidth]{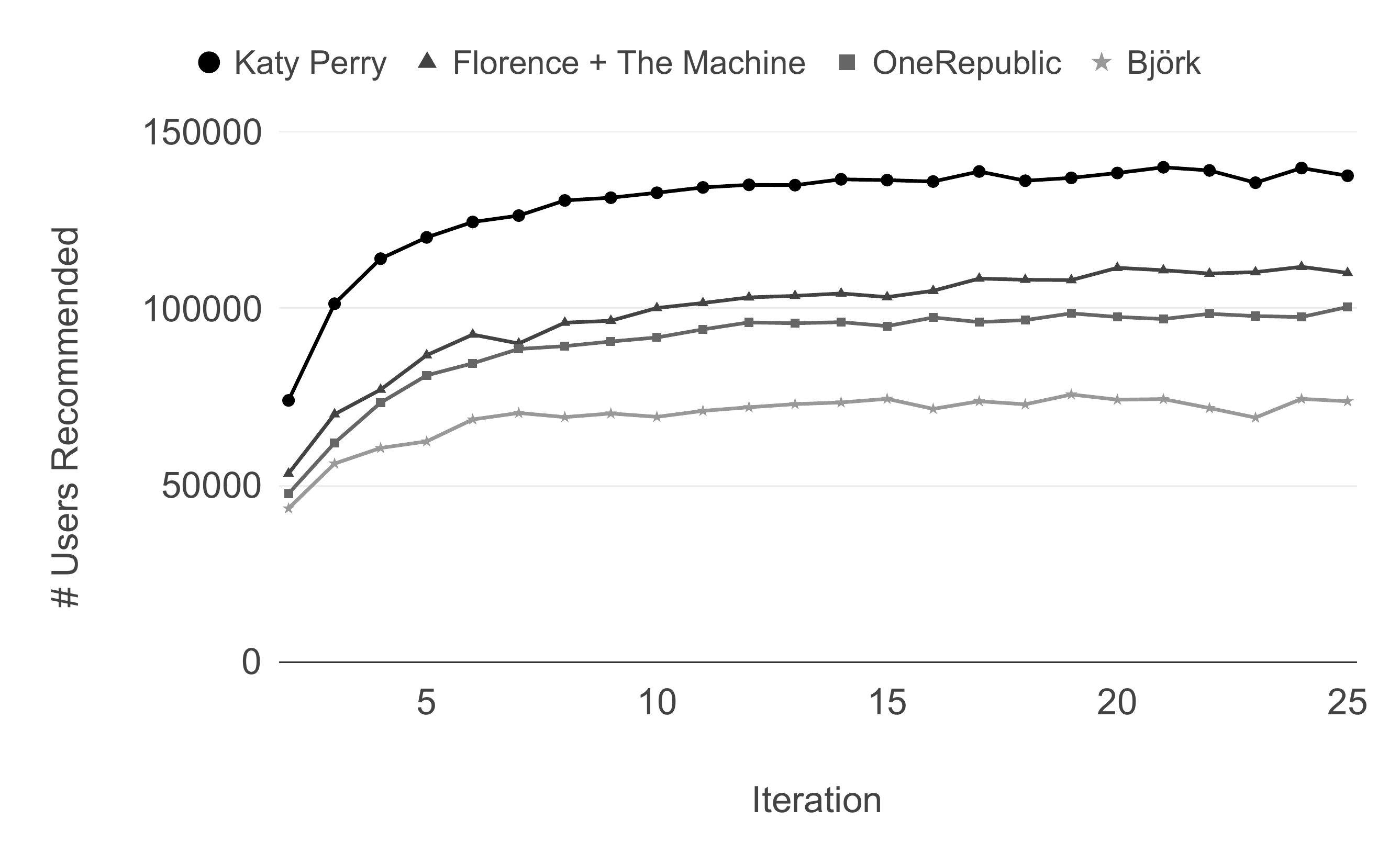}}
 \caption{Number of users reached by a song recommendation on the example of four popular songs when simulating the feedback loop.}
 \label{fig:top}
\end{figure}

In Figure~\ref{fig:top} we demonstrate how the four most played songs according to our initial user-track matrix gather even more exposure from recommendations during the feedback loop iterations. 
These songs have been recommended to between 50,000 and 100,000 users at the first iteration, and ended up being recommended to 100,000 to 135,000 users after 10 iterations, while originally they were listened by around 50,000 users. 

It is important to mention that in a real case there will be other interactions between users and items that are not considered here. 


\section{Conclusions}\label{conclusions}
In this work, we have considered how the popularity bias is affecting collaborative filtering recommendations based on Matrix Factorization. 
In our experiments, this algorithm is increasing the exposure of more popular musical styles, while reducing the exposure in the long tail, which may be an undesired behaviour.

The goal of our future research is to 
expand our analysis on 
state-of-the-art algorithms proposed for 
cold-start and long-tail music recommendation, which are still lacking such an evaluation. 

\section{Acknowledgements}
This research has been supported by Kakao Corp.

\footnotesize
\bibliography{ISMIRtemplate}

%
%
%
%

\end{document}